\documentclass[numberedappendix]{emulateapj}
\usepackage{amsmath}
\usepackage{graphicx}
\usepackage{multirow}
\usepackage{color}
\usepackage{latexsym}
\usepackage{amssymb}
\usepackage{epsfig}

\begin{document}
\title{Light Curves from Supernova Shock Breakout through an Extended Wind}

\author{Sivan Ginzburg and Shmuel Balberg}

\affil{Racah Institute of Physics, The Hebrew University, Jerusalem 91904, Israel}

\begin{abstract}
Recent observations suggest that some supernovae may be the result of an explosion into an optically thick circumstellar material, the product of pre-explosion mass-loss (wind) by the progenitor star. This scenario has been studied previously both analytically and numerically. However, many previous studies base their analysis on the diffusion approximation for radiation transfer, which is inappropriate in the optically thin outer layers of the wind. Here we study the deviations from diffusion, and calculate light curves more accurately using a Monte Carlo approach to photon transfer. We distinguish between ``compact'' winds, for which the diffusion approximation is appropriate, and ``extended'' winds, which require a more delicate treatment of the radiation. We show that this effect is more significant than that of the light travel time difference to a distant observer, which has a secondary influence on the light curves of extended-wind systems. We also comment on the applicability of the widely used flux-limited diffusion approximation in this context: we find that it generally does not reproduce the Monte Carlo results. The flux-limited diffusion approximation leads to results which are not only quantitatively, but also qualitatively wrong, in the extended-wind regime.
\end{abstract}

\keywords{circumstellar matter --- radiative transfer --- shock waves --- stars: mass-loss --- supernovae: general}

\section{Introduction}
\label{sec:Introduction}

A supernova explosion generates a shock wave which propagates through the envelope of the progenitor star. When the shock wave reaches the outer envelope region which has a low optical depth, its energy can escape by radiation to the surface, and the shock dissolves. The emergence of the shock is referred to as ``breakout'', and it is the first electromagnetic indication of the explosion \citep{Arnett1996}. 

The shock breakout through the surface of the star gives rise to a radiation burst which is typically very short in comparison with the main observed light curve. The situation is different in the presence of an optically thick circumstellar material which engulfs the progenitor star. In this case, the breakout timescale is longer, and the entire observed light curve is dominated by the breakout process \citep[see, e.g,][]{Ofek2010,ChevalierIrwin2011,Moriya2011}.

Numerical studies of supernova explosions into circumstellar material have been carried out by \citet{Grassberg1971,FalkArnett1973,FalkArnett1977,GrasbergNadezhin87}, and more recently by \citet{Moriya2011,Moriya2012}. The scenario was also studied analytically, under certain simplifying assumptions \citep{Chevalier1982,ChevalierFransson1994,Balberg2011,ChevalierIrwin2011, Chatzopoulos2012,Chatzopoulos2013,MoriyaTominaga2012}. In a previous work \citep[Ginzburg \& Balberg 2012; hereafter][]{GB2012}, we related the circumstellar material parameters to the resulting light curve, for the particular case of circumstellar material with $r^{-2}$ density profile, the result of a steady mass-loss (wind). We used a simple hydrodynamic diffusion code, in conjunction with analytical limits, to conduct a systematic parameter survey of the problem \citep[see also][]{Moriya2011}. We showed that the shape of the light curve, and especially the typical timescale and the total energy, can serve as quantitative indication of the nature of the progenitor system, including the wind parameters. 

The simplified treatment of radiation in \citet{GB2012}, together with the fit to analytical limits, provide intuition in comparison with more elaborate numerical codes, while the self-consistent coupling of hydrodynamics and radiation adds accuracy to previous analytical studies. However, the diffusion approximation, which was also used as the basis of other previous analytical works, breaks down at low optical depths, which are relevant in shock breakout phenomena.

In the current work, we study the deviations from the diffusion approximation in the wind-engulfed supernovae scenario. We do this by means of a stationary Monte Carlo ``toy model'', which covers the essence of the problem when the expansion velocities are relatively low---which is the case for stellar explosions into massive winds. This analysis serves as a preliminary assessment of the non-diffusive nature of radiation emission; a rigorous description of the shock breakout requires the coupling of hydrodynamics with radiation transport or Monte Carlo methods, which we defer to future work. We use this simpler model to gain insight to the problem while still producing quantitatively significant results.

As in \citet{GB2012}, we focus on  $r^{-2}$ wind density profiles. We do not necessarily advocate for this scenario in terms of the late stages of stellar evolution, but rather use this case as a well defined example which can be studied thoroughly, and offers a simple, realistic context to study a configuration in which the diffusion approximation breaks down. The extension of the results to different wind profiles is straightforward \citep[see][for a discussion of wind shells with different power laws]{MoriyaTominaga2012}.

The outline of the paper is as follows. In Section \ref{sec:progenitor} we formulate our progenitor model, and in Section \ref{sec:timescale} we review the relations between the system parameters and the observed features of the light curve in the diffusion approximation. The Monte Carlo method and the deviations from diffusion are presented in Sections \ref{sec:deviations} and \ref{sec:light_curves}. Corrections due to the effects of light travel to a distant observer are considered in Section \ref{sec:observer}. A summary of our main conclusions is presented in Section \ref{sec:conclusions}. The flux-limited diffusion approximation, which is an alternative method for low optical depth, is discussed in the Appendix.

\section{Progenitor Systems}
\label{sec:progenitor} 

The progenitor system includes two components: the gravitationally bound star and the outer material which is created by pre-explosion mass-loss. The explosion is treated as an instantaneous release of thermal energy $E$ at the center of the star. The star and the wind are approximated as initially stationary and cold. Our models for both components are described in detail in \citet{GB2012}, and we repeat them briefly here.

For the star we assume a single fixed model of a polytrope with radius $R_*=10^{13}$ cm, mass $M_*=15M_\sun$, and polytropic index $n=3/2$. We emphasize that in the progenitor systems discussed here the star is essentially a point mass compared to the size of the wind. For this reason, and due to our interest in systems with wind masses comparable to the ejecta mass $M_w\sim M_{\rm ej}$, the details of its structure are generally unimportant for a given $M_*$.

The circumstellar material is assumed to be the product of a steady wind with mass-loss rate $\dot{M}$ and wind velocity $v_w$, resulting in a density profile of
\begin{equation}\label{eq:wind_density}
\rho(r)=\frac{\dot{M}}{4\pi r^2v_w}\equiv Kr^{-2}.
\end{equation}

For the wind to have a finite mass, the $\rho(r)\propto r^{-2}$ density profile cannot extend to infinity. Here we assume that the wind stretches out from the surface of the star $R_*$ and cuts off at some outer radius $R_w$. For $R_w\gg R_*$ the total wind mass is approximated by
\begin{equation}\label{eq:wind_mass}
M_w\approx 4\pi KR_w.
\end{equation}

An explosion model is therefore set by three parameters, after arbitrarily fixing $M_*=15M_\sun$ for simplicity: the explosion energy $E$, the wind outer radius $R_w$, and the wind density coefficient $K$ from Equation \eqref{eq:wind_density}.

\section{Luminosity and Timescale in the Diffusion Approximation}
\label{sec:timescale}

The main features of the observed light curve are the total radiated energy, the peak luminosity and the characteristic timescale, which are related to each other through the emission of thermal energy at shock breakout. In \citet{GB2012} we derived the qualitative and quantitative relations between these features and the wind parameters when using the diffusion approximation. In this section we repeat the derivation of these relations in brief, and emphasize the limits of their validity.

The total emitted energy, $E_{\rm rad}$, can be estimated through a plastic collision interpretation, giving the simple relation  
\begin{equation}\label{eq:e_rad}
E_{\textrm{rad}}\propto E\frac{M_w}{M_*+M_w}.
\end{equation}

The light curve results when this energy is emitted as the shock breaks out of the wind. Essentially, breakout occurs when the timescale for photons to diffuse from the shocked region to the photosphere is short enough to be comparable to the dynamical time of the shock. Quantitatively, the condition for breakout is 
$\tau \sim c/v_{\rm sh}$, where $\tau$ is the optical depth from the shock to the edge of the wind, $v_{\rm sh}$ is the shock velocity, and $c$ the speed of light \citep{Weaver1976}. The radial position where this condition is fulfilled is defined as the breakout radius, $R_{\rm sh}$. Assuming a constant opacity, $\kappa$, in the $r^{-2}$ density profile of Equation \eqref{eq:wind_density}, the optical depth from any radius $r$ to the wind edge is given by 
\begin{equation}\label{eq:optical_depth}
\tau(r)=\int_r^{R_w}{\kappa\rho dr'}=\int_r^{R_w}{\kappa Kr'^{-2}dr'}=\kappa K\left(\frac{1}{r}-\frac{1}{R_w}\right).
\end{equation}
Using Equation \eqref{eq:optical_depth}, the breakout radius can then be expressed as \citep{ChevalierIrwin2011}
\begin{equation}\label{eq:breakout_rd}
\frac{1}{R_{\rm sh}}\approx\frac{1}{R_w}+\frac{1}{R_d},
\end{equation}
where
\begin{equation}\label{eq:rd}
R_d\equiv\frac{\kappa K v_{\rm sh}}{c}\,
\end{equation}   
is the effective diffusion radius. The physical interpretation of $R_d$ is the shock breakout radius for an infinite $\rho(r)=Kr^{-2}$ wind.

If the radiation flow is diffusive, the relevant timescale of the light curve is the diffusion time from $R_{\rm sh}$ to $R_w$. This time can be estimated by
\begin{equation}\label{eq:diff_integral}
t_d\approx\int_{R_{\rm sh}}^{R_w}{\frac{d(r-R_{\rm sh})^2}{D(r)}}=
\int_{R_{\rm sh}}^{R_w}{\frac{2(r-R_{\rm sh})dr}{D(r)}},
\end{equation}
where the density dependent diffusion coefficient is
\begin{equation}\label{eq:density_d}
D(r)=\frac{c}{3\kappa\rho(r)}.
\end{equation}
Combining Equations \eqref{eq:diff_integral} and \eqref{eq:density_d} with the density profile of Equation \eqref{eq:wind_density}, and with Equation \eqref{eq:breakout_rd}, yields the diffusion time for any $R_w$
\begin{equation}\label{eq:tdiff_rw_rd}
t_d\sim\frac{\kappa K}{c}\left[\ln\left(1+\frac{R_w}{R_d}\right)+\frac{1}{1+R_w/R_d}-1\right].
\end{equation} 

\subsection{Compact and Extended Winds}
\label{subsec:compact_extended}

The key distinction in the progenitor system is the relation between $R_w$ and $R_d$. Equation \eqref{eq:tdiff_rw_rd} can be approximated in both limits of $R_w/R_d$:
\begin{equation}\label{eq:tdiff_limits}
t_d\sim\frac{\kappa K}{c}\cdot
\begin{cases}
\frac{1}{2}\left(\frac{R_w}{R_d}\right)^2 & R_w\ll R_d
\\[1.5ex]
\ln\frac{R_w}{R_d}-1 & R_w\gg R_d
\end{cases}.
\end{equation}
In systems where $R_w\ll R_d$, the breakout occurs when the shock reaches the edge of the wind, since, using Equation \eqref{eq:breakout_rd}, $R_{\rm sh}\approx R_w$. In this limit, diffusion takes place over a small distance compared with $R_w$
\begin{equation}\label{eq:delta_r}
\Delta R=R_w-R_{\rm sh}\approx\frac{R_w^2}{R_d},
\end{equation}
over which the diffusion coefficient is nearly constant 
\begin{equation}\label{eq:const_d}
D=\frac{cR_w^2}{3\kappa K},
\end{equation} 
so that
\begin{equation}\label{eq:const_d_tdiff}
t_d\sim\frac{\Delta R^2}{D}\sim\frac{\kappa K}{c}\left(\frac{R_w}{R_d}\right)^2,
\end{equation}
in compliance with Equation \eqref{eq:tdiff_limits}.

The qualitative discussion and the calculations in \citet{GB2012} focused on reconstructing observed light curves in superluminous supernovae observations. Both limits of Equation \eqref{eq:tdiff_limits} were discussed in \citet{GB2012}, and were found to comply with the full hydro-radiation simulations. However, both the derivation of Equation \eqref{eq:tdiff_limits}, and the treatment of radiation in those simulations are limited by the validity of the diffusion approach. Diffusion is valid as an approximation to radiation transport only at opaque regions, i.e., at high optical depth from the surface \citep[see, e.g.,][]{Zeldovich}. Roughly speaking, the diffusion approximation is valid up to $\tau\sim 1$ (the photosphere), and not all the way to the surface. The photosphere, using Equation \eqref{eq:optical_depth}, is located at $R_{\rm ph}$ which is given by
\begin{equation}\label{eq:rph}
\frac{1}{R_{\rm ph}}\approx\frac{1}{R_w}+\frac{1}{\kappa K}.
\end{equation} 

We make a second distinction between wind profiles with respect to the relation of $R_w$ to the length scale $\kappa K$. We hereafter refer to wind profiles which satisfy $R_w\ll\kappa K$ as ``compact''. Using Equation \eqref{eq:rph}, the photosphere and the wind edge in these winds coincide ($R_{\rm ph}\approx R_w$), so that the photon travel time from the photosphere to the surface is a small correction. Such winds may therefore be treated accurately with the diffusion approximation. Correspondingly, we expect the diffusion approximation to break down for wind profiles in which $R_w\gtrsim\kappa K$. We refer to these winds as ``extended''.  As we argued in \citet{GB2012}, a rough correction to the timescale in this case can be obtained by considering diffusion only to the photosphere, and neglecting the photon transfer from the photosphere to the wind edge. By replacing $R_w$ with $R_{\rm ph}$ in Equation \eqref{eq:tdiff_rw_rd}, we obtain
\begin{equation}\label{eq:diff_to_photosphere}
t_d\sim\frac{\kappa K}{c}\left[\ln\left(\frac{c}{v_{\rm sh}}\frac{1}{1+\left[\kappa K/R_w\right]}\right)-1\right],
\end{equation}
for $R_w\gg R_d$. Equation \eqref{eq:diff_to_photosphere} reduces to Equation \eqref{eq:tdiff_limits} for compact winds, but predicts a constant asymptotic timescale
\begin{equation}\label{eq:asymptote}
t_d\to\frac{\kappa K}{c}\left[\ln\left(\frac{c}{v_{\rm sh}}\right)-1\right],
\end{equation}
for very extended winds ($R_w\gg\kappa K$), which is different from the estimate of Equation \eqref{eq:tdiff_limits}. For a similar analytical treatment of the optically thin $\tau\lesssim 1$ region, see \citet{MoriyaTominaga2012}. 

The wind models for superluminous supernovae, presented in \citet{GB2012}, correspond to the regime $R_w\sim R_d=\kappa Kv_{\rm sh}/c\ll\kappa K$ (for non-relativistic radiation mediated shocks which we consider here, $v_{\rm sh}\ll c$, and therefore $R_d\ll\kappa K$). Consequently, these models are categorized as compact, and the approximated diffusion treatment of radiation in \citet{GB2012} and previous works \citep{Ofek2010,ChevalierIrwin2011} is justified.

\section{Quantitative Analysis of the Deviation from Diffusion in Extended Winds}
\label{sec:deviations}

We now turn to a more careful analysis of the deviation from the diffusion approximation in the extended wind regime. We estimate this deviation by comparing calculated light curves of a system with given initial conditions. The light curves are calculated both in the diffusion approximation, and using a Monte Carlo based code. 

We begin by redefining the problem in a dimensionless form. Motivated by Section \ref{subsec:compact_extended}, we use $\kappa K$ as a characteristic length scale. For example, the dimensionless wind radius, $\tilde R_w$, is defined by  $R_w=\kappa K\tilde R_w$. Using this notation, Equation \eqref{eq:breakout_rd} for the shock breakout radius can be written as
\begin{equation}\label{eq:rsh_nodim}
\frac{1}{\tilde R_{\rm sh}}=\frac{1}{\tilde R_w}+\frac{1}{\tilde R_d},
\end{equation} 
with
\begin{equation}\label{eq:rd_nodim}
\tilde R_d=\frac{v_{\rm sh}}{c}.
\end{equation}
Similarly, using Equation \eqref{eq:rph}, the photosphere is defined by
\begin{equation}\label{eq:rph_nodim}
\frac{1}{\tilde R_{\rm ph}}=\frac{1}{\tilde R_w}+1.
\end{equation}
The compact wind condition in this notation is $\tilde R_w\ll 1$, and our interest will be in cases where $\tilde R_w\gtrsim 1$. In this dimensionless formulation, we measure time with the characteristic timescale $\kappa K/c$.
 
In Section \ref{subsec:stationary} we present the test problem used to evaluate the deviation from diffusion. The numerical methods (diffusion and Monte Carlo) are described in Sections \ref{subsec:diffusion} and \ref{subsec:monte}. The resulting light curves are presented in Section \ref{sec:light_curves}.

\subsection{The Stationary Model}
\label{subsec:stationary}

Our main approximation in this work is replacing the hydrodynamical description of the supernova explosion with a stationary model of the shock breakout. By doing so we obviously lose some of the fundamental features of the problem. Most notably, we neglect the expansion of the gas with time, and we treat the interaction of the expanding ejecta and wind as an instantaneous release of energy. On the other hand, the stationary model reduces the problem essentially to a single parameter---$\tilde R_w$ (since $\tilde R_d$ is limited to a relatively narrow range around $\sim 10^{-2}$ for non-relativistic shock breakouts through a wind). This allows for a generalized analysis, which captures the main physics of the problem of breakout through an extended wind, for which the diffusion approximation becomes inaccurate. Moreover, in supernovae which interact with a massive wind (and especially the SLSNe limit), the shock and expansion velocities are low, so the photon transfer (either diffusion or free streaming) timescale is shorter than the timescale over which the position and overall structure of the region through which the photons travel change, and therefore a stationary approximation is not unreasonable. A concrete example may be found in \citet{GB2012}, Figures 2-5: in this example the velocities in front of the shock are low enough for the region through which photons diffuse to remain approximately stationary during the emission of radiation. Also, note that the qualitative estimates of Section \ref{sec:timescale} and \citet{GB2012}, which fit the dynamic hydro-diffusion calculations well, are also based on the assumption of stationarity. 

Following the discussion in Section \ref{sec:timescale}, for each choice of $\tilde R_w$ and $\tilde R_d$ we set a stationary configuration with an inner radius $\tilde R_{\rm sh}$, given by Equation \eqref{eq:rsh_nodim}, and an outer radius $\tilde R_w$. Note that treating $\tilde R_d=v_{\rm sh}/c$ as a constant, independent of the value of $\tilde R_w$, is also an approximation, justified by the fact that $v_{\rm sh}$ is approximately constant for both $R_w\ll R_d$ and $R_w\gg R_d$ limits \citep[see][]{GB2012}. We fix $\tilde R_d=10^{-2}$, motivated by order of magnitude estimates and the hydro-diffusion calculations in \citet{GB2012}.

\begin{figure}[tbh]
\epsscale{1} \plotone{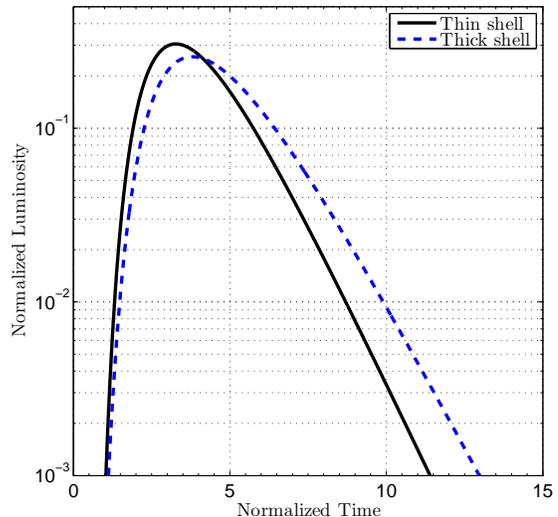}
\caption{Calculated normalized (see text) light curves for $\tilde R_d=10^{-2}$ and $\tilde R_w=8\times 10^{-1}$. The light curves were obtained using diffusion calculations. The calculations are of an infinitesimally thin hot shell, situated at $\tilde R_{\rm sh}$ (solid black line) and of a thick hot shell, which extends from $\tilde R_{\rm sh}$ to $\tau=10^3$ from the surface (dashed blue line).
\label{fig:light_curves_shell}}
\end{figure}

We generate the light curve in a given system by calculating the photon flux at the outer radius of the simulation. In order to simplify the calculations and remove the last free parameter, we release all the photons at $\tilde R_{\rm sh}$, allowing them to propagate outward. Injecting the energy into an infinitesimally thin hot shell, is another deviation from the realistic hydrodynamic problem, in which the radiating shell has a finite width; for example, we refer to Figure 2 of \citet{GB2012}, which shows that, at breakout, the hot shell extends from $\tilde R_{\rm sh}$ to an optical depth of roughly $10^3$. While taking into account the finite width of the shell must widen the light curve, this is a minor effect since the hot shell is approximately isothermal, and so the bulk of the radiation energy is indeed concentrated at the outer edge of the hot shell, near $\tilde R_{\rm sh}$. We demonstrate this effect by comparing in Figure \ref{fig:light_curves_shell} two diffusion calculations (see Sections \ref{subsec:diffusion} and \ref{sec:light_curves} for details), one with a thin radiating shell (our standard simplified model), and one with a more physical, thick isothermal shell, which extends to $\tau=10^3$. We find that the effect of this distinction is at a level of roughly 15\% in the timescale and peak luminosity, and does not change the light curve qualitatively. Since we do not presume to calculate exact supernovae light curves in the current work, but only to estimate the validity of the diffusion approximation by comparing it to Monte Carlo calculations, we proceed with the simple, well defined, thin shell stationary model. 

\subsection{The Diffusion Method}
\label{subsec:diffusion}

We calculate the light curve in the diffusion approximation by solving the diffusion equation for the radiation energy density $U$
\begin{equation}\label{eq:diffusion_equation}
\frac{\partial U}{\partial t}=\nabla\left(D\nabla U\right),
\end{equation}
with the diffusion coefficient $D$ given by Equation \eqref{eq:density_d}. The calculations are performed using dimensionless quantities, and applied to the stationary model, presented in Section \ref{subsec:stationary}. The dimensionless version of Equation \eqref{eq:diffusion_equation} is solved implicitly, using a tridiagonal system of equations. Precision is obtained by limiting the time step to allow only for small changes in $\tilde U$ at each step. The configuration is divided to 800 cells, geometrically increasing in size to obtain sufficient resolution both at the opaque and transparent regions. The initial condition we set is a unit energy released at the innermost cell, adjacent to the closed inner boundary. The outer boundary is an open one. 

\subsection{The Monte Carlo Method}
\label{subsec:monte}

There are several methods for solving radiation transfer more appropriately in transparent, low optical depth, regions. The most straightforward techniques are Monte Carlo methods which sample the photon distribution with a finite number of particles, which undergo events (such as scattering, absorption, etc.) according to the appropriate probability. In principle, Monte Carlo methods can include all the relevant physical processes, and can therefore be exact both in opaque and transparent regions, including the transition. The main drawback of these methods is their computational cost. Nevertheless, a simple Monte Carlo algorithm, described below, is sufficient for an estimate of the deviations from diffusion in our scenario.

In our calculation each photon is initialized at the inner radius $\tilde r=\tilde R_{\rm sh}$. From there, the photon is advanced in successive Monte Carlo steps which are detailed below, while keeping track of the photon radius $\tilde r$.
First, a random direction of motion $\theta\in[0,\pi]$, relative to the current radius vector, is chosen (in three dimensions, $\cos\theta$ is uniformly distributed). Next, we choose a random optical distance $\Delta\tau$ for the step, distributed by $e^{-\Delta\tau}$. In order to find the location at the end of the step, we calculate the geometrical distance, $\tilde s(\Delta\tau,\theta,\tilde r_0)$, which is a function of the optical distance, the direction, and the radius at the end of the previous step $\tilde r_0$. The relation between the optical distance and the geometrical distance is given by (we revert temporarily to physical, not dimensionless, quantities)
\begin{equation}\label{eq:int_free_path}
\Delta\tau=\int_0^s{\frac{dx}{l(x)}}=\int_0^s{\kappa\rho(x)dx},
\end{equation}
with $l=1/\kappa\rho$ is the mean free path. For $r^{-2}$ density profile
\begin{equation}\label{eq:int_delta_tau}
\Delta\tau=\kappa K\int_0^s{\frac{dx}{r^2}}=\kappa K\int_0^s{\frac{dx}{r_0^2+x^2+2r_0x\cos\theta}}.
\end{equation}
The integration of Equation \eqref{eq:int_delta_tau} yields
\begin{equation}\label{eq:delta_tau_integrated}
\Delta\tau=\frac{\kappa K}{r_0\sin\theta}\left[\arctan\left(\frac{s}{r_0\sin\theta}+\cot\theta\right)+\theta-\frac{\pi}{2}\right].
\end{equation}
We rewrite Equation \eqref{eq:delta_tau_integrated} with the dimensionless notation (lengths measured in $\kappa K$)
\begin{equation}\label{eq:delta_tau}
\Delta\tau=\frac{1}{\tilde r_0\sin\theta}\left[\arctan\left(\frac{\tilde s}{\tilde r_0\sin\theta}+\cot\theta\right)+\theta-\frac{\pi}{2}\right].
\end{equation} 
Inverting the relation in Equation \eqref{eq:delta_tau}, we get the geometrical distance
\begin{equation}\label{eq:geo_distance}
\tilde s=\tilde r_0\sin\theta\left[\cot(\theta-\tilde r_0\sin\theta\Delta\tau)-\cot\theta\right].
\end{equation}
Equation \eqref{eq:geo_distance} is physical only as long as $\theta-\tilde r_0\sin\theta\Delta\tau$ is positive, corresponding to a finite travel distance. Due to the $r^{-2}$ dependence of the density profile, for each radius and direction of propagation, the optical depth from the point of origin to infinity is finite (unless the path is directed exactly toward the origin). This corresponds to a finite critical optical distance $\Delta\tau_c$, which is the solution of the equation
\begin{equation}\label{eq:crit_opt}
\theta-\tilde r_0\sin\theta\Delta\tau_c=0.
\end{equation}
For $\Delta\tau>\Delta\tau_c$, the photon escapes to infinity.

After calculating the geometrical distance $\tilde s$ that the photon travels, we calculate the radius of its new position at the end of the step
\begin{equation}\label{eq:new_rad}
\tilde r^2=\tilde r_0^2+\tilde s^2+2\tilde r_0\tilde s\cos\theta.
\end{equation} 
Next, one of three options is executed:\\
{\it Escape:} if the new radius is larger than the outer radius ($\tilde r>\tilde R_w$), or if the photon escapes to infinity, the simulation stops, and the final distance to the edge,
\begin{equation}\label{eq:dist_to_edge}
\tilde s_f=\sqrt{\tilde R_w^2-(\tilde r_0\sin\theta)^2}-\tilde r_0\cos\theta,
\end{equation}
is added to the accumulated travel distance.\\
{\it Reflection from the inner boundary:} if the new radius is smaller than the inner radius ($\tilde r<\tilde R_{\rm sh}$), then we set the photon to begin the next step from $\tilde R_{\rm sh}$ and add the distance
\begin{equation}\label{eq:dist_to_inner}
\tilde s_i=-\sqrt{\tilde R_{\rm sh}^2-(\tilde r_0\sin\theta)^2}-\tilde r_0\cos\theta,
\end{equation}
to the total travel distance. This choice is equivalent to the closed inner boundary condition in the diffusion approximation.\\
{\it Further scattering:} finally, if the new radius is within limits ($\tilde R_{\rm sh}<\tilde r<\tilde R_w$), then we simply add the distance $\tilde s$ to the accumulated travel distance and continue to another Monte Carlo step.

Each photon eventually exits $\tilde R_w$ with some accumulated travel distance, which, through the choice of $c=1$, is equivalent to its travel time. 
By sampling enough photons (we use $10^7$ for each calculation), and dividing their travel times into bins, we construct a light curve.

\section{Light Curves}
\label{sec:light_curves}

We now present the main results of the numerical calculations, conducted by both methods described in Sections \ref{subsec:diffusion} and \ref{subsec:monte}. The results are presented with distances normalized to $\kappa K$, times to $\kappa K/c$, and the total radiated energy in each light curve, $E_{\rm rad}$, is 1, unless stated otherwise.

\subsection{Diffusion and Monte Carlo Calculations in the Stationary Model}
\label{subsec:diff_monte_stat}

\begin{figure}[tbh]
\epsscale{1} \plotone{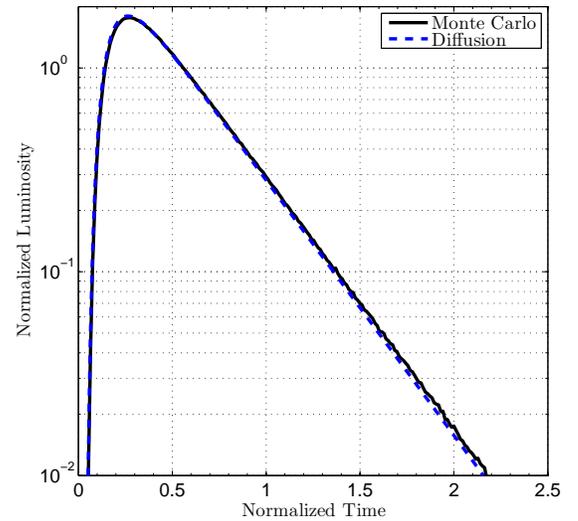}
\caption{Calculated normalized (see text) light curves for $\tilde R_d=10^{-2}$ and $\tilde R_w=1.3\times 10^{-2}$. The plotted light curves were obtained using a Monte Carlo calculation (solid black line) and a diffusion calculation (dashed blue line).
\label{fig:light_curves}}
\end{figure}

In Figure \ref{fig:light_curves} we present calculated light curves for an example compact wind with $\tilde R_w=1.3\times 10^{-2}$. As expected, in this region the diffusion approximation is valid, and therefore complies with the Monte Carlo results: the two light curves are identical. In essence, this comparison serves as a check of our simulations.

We note that the stationary light curves (in both methods) do not include a low luminosity tail at the declining part of the light curve, which is typically found in hydrodynamic calculations. The origin of the tail is the dynamic nature of the problem at long times, especially the ongoing conversion of kinetic energy into thermal energy as the ejecta and the wind continue to interact. Since such an effect is not included in the stationary approximation, the light curves drop after their maximum luminosity without a tail. While we address this phenomena more extensively in Section \ref{subsec:dynamic}, we note now that the tail does not effect key features such as the peak luminosity and the light curve rise time. 

\begin{figure}[tbh]
\epsscale{1} \plotone{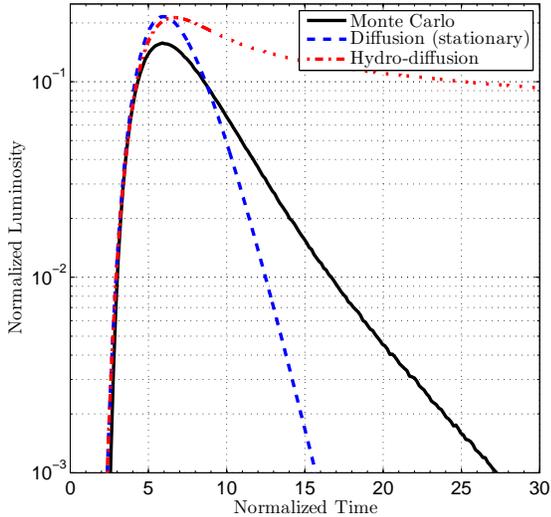}
\caption{Calculated normalized (see text) light curves for $\tilde R_d=10^{-2}$ and $\tilde R_w=1.3\times 10^1$. The plotted light curves were obtained using a Monte Carlo calculation (solid black line) and a diffusion calculation (dashed blue line). The Monte Carlo calculation is shifted in time. For qualitative comparison, a hydro-diffusion calculation (dot-dashed red line) is given for the values $K=10^{16} \textrm{ g cm}^{-1}$, $\kappa=0.34\textrm{ cm}^2\textrm{g}^{-1}$, $E=5\times 10^{51}$ erg, and $R_w=4.4\times 10^{16}$ cm ($\tilde R_w=1.3\times 10^1$). The hydro-diffusion light curve is normalized to $2.8\times 10^{44}$ erg $\rm{s}^{-1}$ and shifted in time.
\label{fig:light_curves_ext}}
\end{figure}

In Figure \ref{fig:light_curves_ext} we present calculated light curves for an example extended wind with $\tilde R_w=1.3\times 10^1$. In this regime we expect the diffusion approximation to be inaccurate, and indeed the diffusion calculations show deviations from the Monte Carlo results. Quantitatively, we see that the Monte Carlo calculated light curve is significantly shallower and broader than its diffusion counterpart. In particular, the peak luminosity in the Monte Carlo calculation is about 30\% lower than the corresponding value in the diffusion calculation, while the full width at half maximum (FWHM) timescale is longer by about 30\%.

\begin{figure}[tbh]
\epsscale{1} \plotone{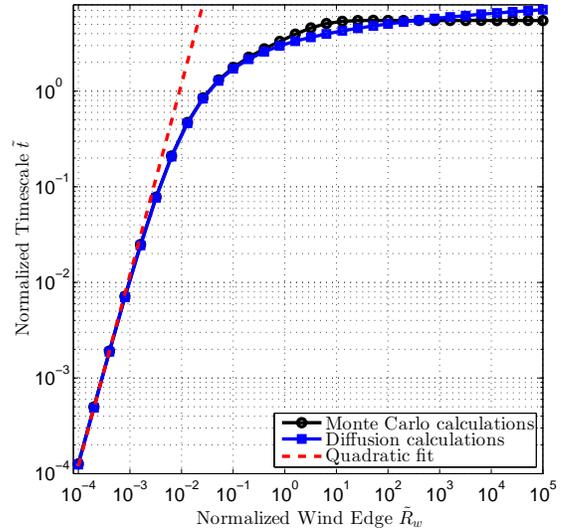}
\caption{Calculated normalized (see text) light curve timescales $\tilde t$ (full width at half maximum) as a function of the normalized wind outer radius $\tilde R_w$. The results are plotted for $\tilde R_d=10^{-2}$. The light curves are calculated using the Monte Carlo method (solid black line, marked with circles) and the diffusion method (solid blue line, marked with squares). Each marker represents a single simulation. The analytic fit (dashed red line) is quadratic in $\tilde R_w$.
\label{fig:time_scale}}
\end{figure}

\begin{figure}[tbh]
\epsscale{1} \plotone{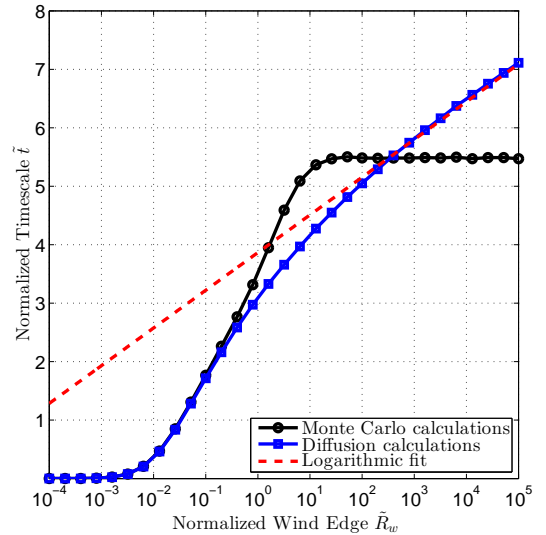}
\caption{Same as Figure \ref{fig:time_scale}, but plotted in a semi-logarithmic scale. The analytic fit (dashed red line) is logarithmic in $\tilde R_w$.
\label{fig:time_scale_semi}}
\end{figure}

We begin by focusing on the FWHM timescale because of its importance in observationally inferring the wind radius \citep{GB2012}. By repeating our calculation with various values of $\tilde R_w$, we conducted a full survey of this timescale as calculated with both methods. The results are shown in Figures \ref{fig:time_scale} and \ref{fig:time_scale_semi}.

As seen in Figures \ref{fig:time_scale} and \ref{fig:time_scale_semi}, the Monte Carlo and diffusion results agree for the compact region ($\tilde R_w\ll 1$), but differ in the extended regime ($\tilde R_w\gtrsim 1$). Figure \ref{fig:time_scale} clarifies that at the extremely compact regime ($\tilde R_w\ll \tilde R_d$), the results follow the homogeneous diffusion estimate of Equations \eqref{eq:tdiff_limits} and \eqref{eq:const_d_tdiff}, which predict a quadratic behavior. The difference between the diffusion and Monte Carlo calculations becomes remarkable in the ultra extended limit ($\tilde R_w\gg 1$) which is displayed in Figure \ref{fig:time_scale_semi}. The diffusion results exhibit a logarithmic growth of the timescale, as predicted by Equation \eqref{eq:tdiff_limits}, while the logarithmic behavior of the Monte Carlo timescales breaks down, leading to a constant asymptotic timescale, in agreement with Equation \eqref{eq:asymptote}. This difference is a result of applying the diffusion approximation to the optically thin $\tau\lesssim 1$ region, which is our main interest here. In this region the photons do not diffuse, but rather ``free stream''---travel without collisions, therefore without widening the light curve. Artificially truncating the diffusion calculation at $\tau\sim 1$, as conducted in Equation \eqref{eq:diff_to_photosphere}, seems to reproduce this behavior, at least qualitatively.

\begin{figure}[tbh]
\epsscale{1} \plotone{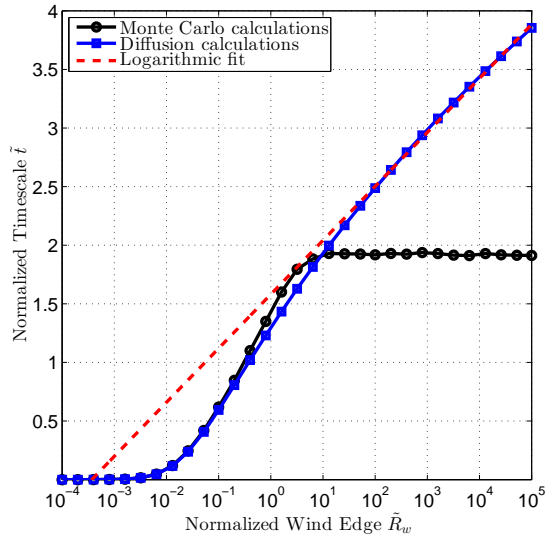}
\caption{Same as Figure \ref{fig:time_scale_semi}, but with normalized timescales $\tilde t$ representing the light curve rise time (from 10\% to 90\% of the peak luminosity) instead of the full width at half maximum. 
\label{fig:time_scale_semi_rise}}
\end{figure}

\begin{figure}[tbh]
\epsscale{1} \plotone{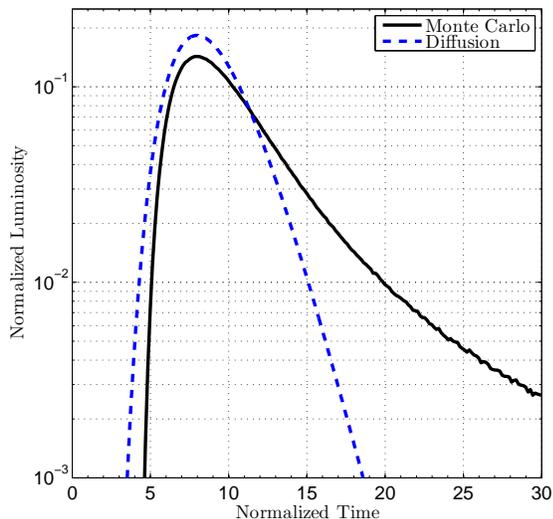}
\caption{Calculated normalized (see text) light curves for $\tilde R_d=10^{-2}$ and $\tilde R_w=10^2$. The plotted light curves were obtained using a Monte Carlo calculation (solid black line) and a diffusion calculation (dashed blue line). The Monte Carlo calculation is shifted in time to match peak luminosities.
\label{fig:light_curves_ext_rise}}
\end{figure}

We note that in many cases, the light curve rise time, rather than the FWHM, is the interesting timescale, especially from an observational perspective. In Figure \ref{fig:time_scale_semi_rise} we repeat the survey, presented in Figures \ref{fig:time_scale} and \ref{fig:time_scale_semi}, with the timescales representing the light curve rise time (from 10\% to 90\% of the peak luminosity) instead of the full width at half maximum. As seen by comparing Figure \ref{fig:time_scale_semi_rise} to Figure \ref{fig:time_scale_semi}, the rise time behaves qualitatively similar to the FWHM, but with some quantitative difference. For example, as seen in Figure \ref{fig:time_scale_semi_rise}, the light curves presented in Figure \ref{fig:light_curves_ext} (with $\tilde R_w=1.3\times 10^1$) have different FWHM but similar rise times. Motivated by Figure \ref{fig:time_scale_semi_rise}, we present in Figure \ref{fig:light_curves_ext_rise} light curves for $\tilde R_w=10^2$, which differ by about 30\% in their rise times.  

\subsection{Comparison with Dynamic Simulations}
\label{subsec:dynamic}

\begin{figure}[tbh]
\epsscale{1} \plotone{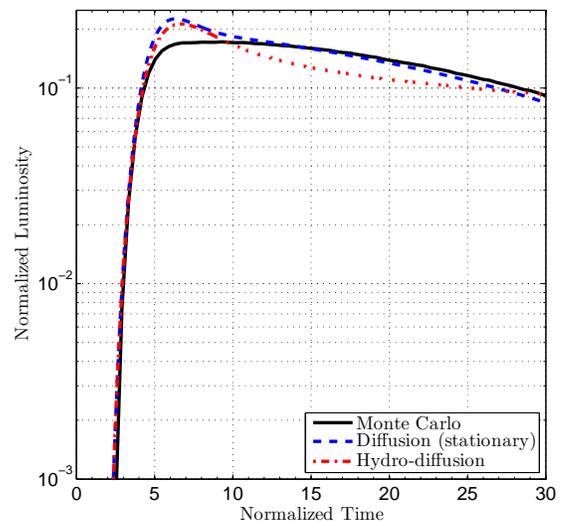}
\caption{Same as Figure \ref{fig:light_curves_ext}, but with an additional energy source term (see text) $\dot E_{\rm rad}(t)=a-bt$ added to the stationary calculations beginning from time $t_{\rm initial}$. The additional energy term parameters are a=0.2, b=0.005, and $t_{\rm initial}=2$. 
\label{fig:light_curves_ext_edot}}
\end{figure}

For qualitative comparison, we add a normalized light curve calculated by the hydro-diffusion code, described in \citet{GB2012}, for a wind mass of $2.8M_\sun$. As seen in Figure \ref{fig:light_curves_ext}, the hydrodynamic and stationary light curves ascend to maximum luminosity in a qualitatively similar manner and with similar rise times, with the quantitative difference is the result of $\tilde R_d$ being only similar to, and not exactly, $10^{-2}$ in the hydrodynamic simulation, and of the general inaccuracies of the stationary model (see Section \ref{subsec:stationary}). The difference in the low luminosity tail, in the declining part of the light curve, is, as noted above, due to the dynamic nature of the problem at long times, especially the ongoing conversion of kinetic energy to thermal energy by interaction of the ejecta and wind. The stationary model does not take this effect into account, so we mimic this additional energy injection artificially by adding a time dependent energy source term, $\dot{E}$, located at the inner boundary, of the form
\begin{equation}\label{eq:edot}
\dot E_{\rm rad}(t)=a-bt,
\end{equation} 
beginning at some initial time $t_{\rm initial}$. This additional energy is added to the original $E_{\rm rad}^0=1$. This serves as a qualitative demonstration in the context of evaluating the deviation from the diffusion approximation. The effects of this additional energy injection, added both to the diffusion and to the Monte Carlo stationary calculations (for $\tilde R_w=1.3\times 10^1$), are displayed in Figure \ref{fig:light_curves_ext_edot}. The parameters of the additional energy term of Equation \eqref{eq:edot} in Figure \ref{fig:light_curves_ext_edot} were chosen so that the tail of the stationary diffusion calculation fits the hydro-diffusion light curve tail up to about 30\%.  

\begin{figure}[tbh]
\epsscale{1} \plotone{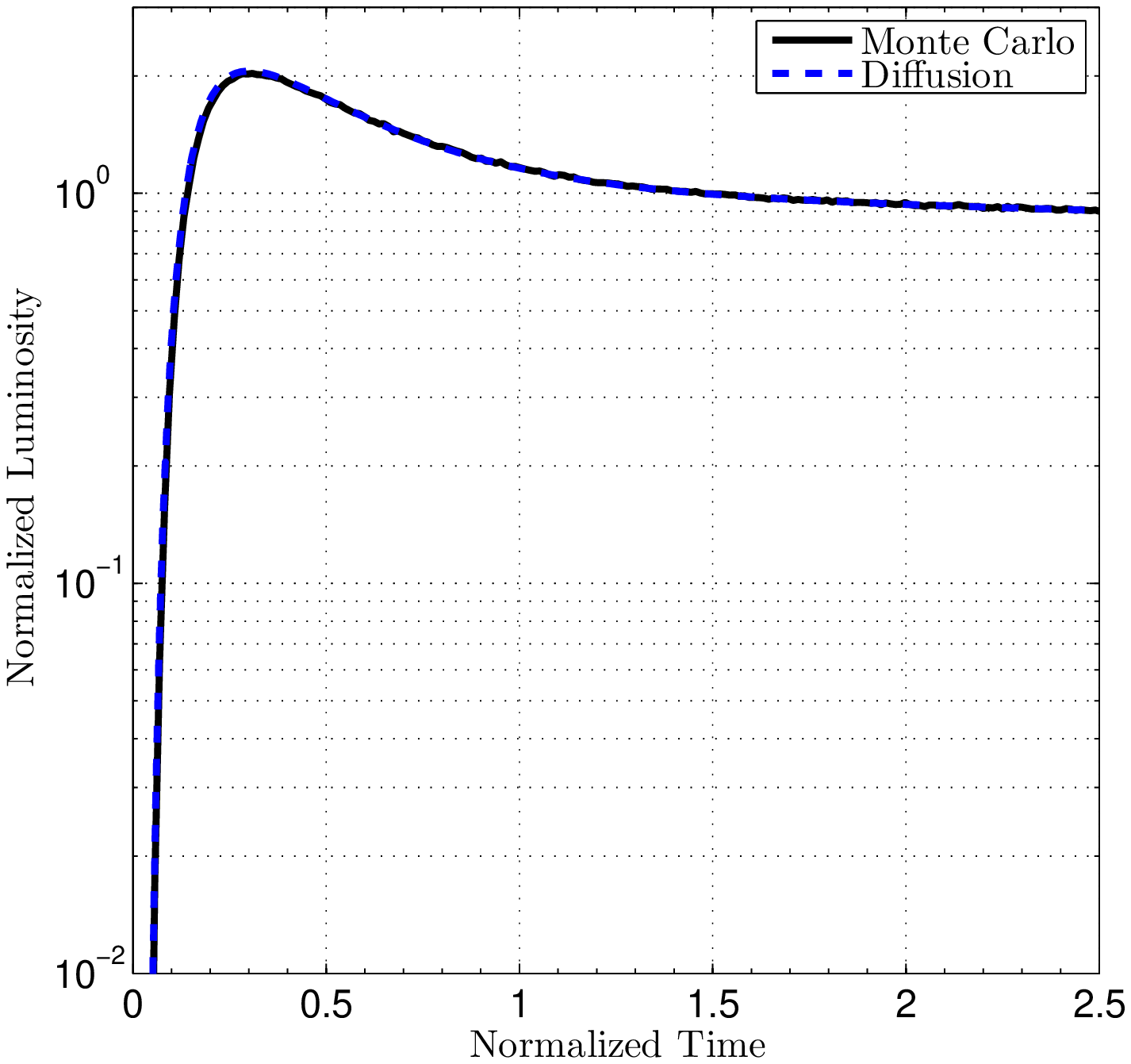}
\caption{Same as Figure \ref{fig:light_curves}, but with an additional energy source term (see text) $\dot E_{\rm rad}(t)=a-bt$ added to the calculations beginning from time $t_{\rm initial}$. The additional energy term parameters are a=1, b=0.05, and $t_{\rm initial}=0$. 
\label{fig:light_curves_edot}}
\end{figure}

As seen in Figure \ref{fig:light_curves_ext_edot}, the low luminosity tail does not have a qualitative effect on the difference between diffusion and Monte Carlo calculations. Quantitatively, the Monte Carlo light curve is still about 30\% wider (in FWHM) and 30\% shallower (at peak luminosity) in comparison with the diffusion calculation. As expected, and as demonstrated in Figure \ref{fig:light_curves_edot} (for $\tilde R_w=1.3\times 10^{-2}$), a similar time dependent energy injection in the compact regime does not change the exact identity between the diffusion and Monte Carlo light curves in this case.   

Although we mimic the dynamic nature of the shock breakout in this section with a simple energy injection, it is noteworthy that other dynamic mechanisms are taken into account in the hydro-diffusion calculations, which might also have an effect on the light curve. For example, the reduction in the optical depth of the circumstellar material in front of the shock, as the shock continues to propagate \citep{Chatzopoulos2013} has potentially a similar effect on the light curve as an energy injection. Since the hydro-diffusion calculation takes this and other effects into account, and since the energy injection recipe in this section produces light curves similar to the dynamic light curves, we assume that these effects are effectively taken into account in the energy injection parameters of Equation \eqref{eq:edot}.     

\section{Light Travel Time to Observer}
\label{sec:observer}

We digress from the main theme of this work to discuss the effect of the light travel time on the observed light curve. This digression is important since light travel time also works to distort the light curve, and hence should be compared to the difference between diffusion and Monte Carlo calculations, in order to asses the importance of accurately modeling the radiation transfer when considering the total bolometric emission.

A distant observer will measure a wider light curve due to the difference in arrival times of photons, which escape from different points on the wind edge shell. The time delay of a photon emitted from some point along the wind edge shell, when compared to the point on the wind edge shell closest to the observer, is
\begin{equation}\label{eq:travel_delay}
\Delta t=\frac{R_w(1-\cos\alpha)}{c},
\end{equation}
where $\alpha$ is the angle between the radius vector and the direction of motion of the photon, at the wind edge.

This time delay may be accounted for by convolution of the original light curve with the distribution of the angle $\alpha$ \citep[see, e.g.,][]{Katz2012}. However, it is both simpler and exact to include this time delay explicitly in the Monte Carlo simulation. The angle $\alpha$ is related to the angle $\theta$ and the radius $\tilde r_0$ of the last scattering before escape (see Section \ref{subsec:monte}) by
\begin{equation}\label{eq:sine_law}
\frac{\tilde r_0}{\sin\alpha}=\frac{\tilde R_w}{\sin\theta}.
\end{equation}
Then ,the normalized time delay,
\begin{equation}\label{eq:travel_delay_nodim}
\Delta\tilde t=\tilde R_w(1-\cos\alpha),
\end{equation}
is added to the accumulated travel time. 

Prior to the presentation of the results, it is important to explain why this time delay could be neglected in previous works \citep{Ofek2010,ChevalierIrwin2011,Moriya2011,GB2012}. In compact winds, the photosphere and the edge of the wind coincide. This means that the last scattering surface is at the edge of the wind, implying that $\tilde r_0\approx\tilde R_w$, and therefore, from Equation \eqref{eq:sine_law}, $\alpha\approx\theta$. 
The distribution $p(\mu)d\mu$ of $\mu\equiv\cos\alpha=\cos\theta$, for the last scattering in this case is \citep{Chandra,Katz2012}
\begin{equation}
p(\mu)d\mu\approx(0.85+1.725\mu)\mu d\mu,
\end{equation}
which has a full width at half maximum (FWHM) of 0.35 in $\cos\alpha$, implying, from Equation \eqref{eq:travel_delay_nodim} a contribution of
\begin{equation}\label{eq:t_obs35}
\tilde t_{\rm obs}=0.35\tilde R_w
\end{equation}
to the light curve timescale. 

For compact winds ($\tilde R_w\ll 1$), we can compare the light travel time difference to the diffusion timescale, given in Section \ref{sec:timescale}. For $\tilde R_d\ll\tilde R_w\ll 1$, the dimensionless version of Equation \eqref{eq:tdiff_limits} gives $\tilde t_d\sim 1$, and therefore the travel time difference to the observer, $\tilde t_{\rm obs}\ll 1$, is negligible. In the opposite limit $\tilde R_w\ll\tilde R_d\ll 1$, Equations \eqref{eq:tdiff_limits} and \eqref{eq:const_d_tdiff} predict that the diffusion time is $\tilde t_d\sim(\tilde R_w/\tilde R_d)^2$. Most previous studies which reconstructed observed superluminous supernovae \citep{ChevalierIrwin2011,Moriya2011,GB2012} dealt with systems which are consistent with the intermediate regime $\tilde R_w\sim\tilde R_d\ll 1$, and therefore are unaffected by light travel time. Note that the travel time difference to the observer does become important, but only for large enough $\tilde R_d$, at least $\tilde R_d^2\gtrsim \tilde R_w$. For $\tilde R_w\ll\tilde R_d^2$, the light curve timescale is actually dominated by the light travel time difference (since $\tilde t_{\rm obs}\gg\tilde t_d$). This limit essentially corresponds to bare stars. 

\begin{figure}[tbh]
\epsscale{1} \plotone{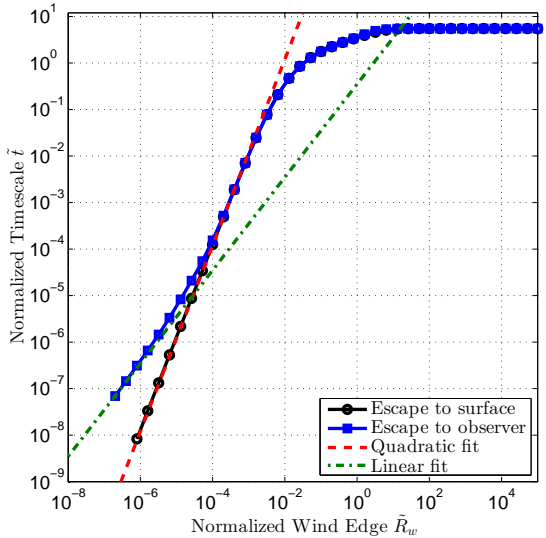}
\caption{Calculated normalized (see text) light curve timescales $\tilde t$ (full width at half maximum) as a function of the normalized wind outer radius $\tilde R_w$. The results are plotted for $\tilde R_d=10^{-2}$. The two plotted sets are with travel time to observer ignored (same as Figure \ref{fig:time_scale})/included (solid black/blue line, marked with circles/squares). Each marker represents a Monte Carlo simulation. The quadratic fit (dashed red line) is the same as in Figure \ref{fig:time_scale}. The linear fit (dot-dashed green line) coefficient is given by Equation \eqref{eq:t_obs35}.
\label{fig:travel}}
\end{figure}

The considerations above do not necessarily hold for extended winds. In general, the light travel time difference, using Equations \eqref{eq:sine_law} and \eqref{eq:travel_delay_nodim} is
\begin{equation}
\Delta\tilde t=\tilde R_w\left[1-\left(1-\frac{\tilde r_0^2}{\tilde R_w^2}\sin^2\theta\right)^{1/2}\right].
\end{equation}
For extended winds, the last scattering surface may be deep inside the wind, at $\tilde r_0\ll\tilde R_w$, and not at $\tilde r_0\approx\tilde R_w$. In this case, the light travel time difference is
\begin{equation}
\Delta\tilde t\approx\frac{1}{2}\tilde R_w\left(\frac{\tilde r_0}{\tilde R_w}\right)^2\sin^2\theta,
\end{equation}
which results in time differences smaller than $\sim\tilde R_w$. For the very extended limit, $\tilde R_w\gg 1$, and the last scattering surface is roughly the photosphere, at $\tilde r_0\sim 1$, so $\Delta\tilde t\ll 1$, and therefore negligible in this limit.

\begin{figure}[tbh]
\epsscale{1} \plotone{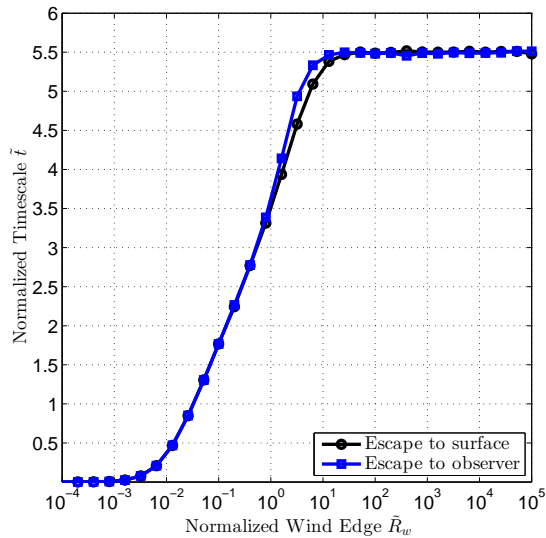}
\caption{Same as Figure \ref{fig:travel}, but plotted in a semi-logarithmic scale.
\label{fig:travel_semi}}
\end{figure}

The Monte Carlo calculations, with travel time to the observer included, are presented in Figures \ref{fig:travel} and \ref{fig:travel_semi}. For compact winds ($\tilde R_w\ll 1$), as seen in Figure \ref{fig:travel}, the travel time to the observer does play a role only for $\tilde R_w\lesssim\tilde R_d^2$, as explained above. This contribution fits well the analytical limit of Equation \eqref{eq:t_obs35}. For extended winds ($\tilde R_w\gtrsim 1$), the travel time to the observer has a minor effect (usually negligible, limited to about 10\% at most) on the light curve timescale, as seen in Figure \ref{fig:travel_semi}. We conclude that for extended winds the light travel time differences are of secondary importance with respect to deviations from the diffusive radiation flow.

\section{Conclusions and Discussion}
\label{sec:conclusions} 

In this work we studied light curves from supernovae exploding in an optically thick extended wind. A wind is considered extended when the optically thin region adjacent to the wind outer edge ($\tau\lesssim 1$) is spatially large enough so that it plays an important role in photon propagation, whereas in a compact wind this optically thin region does not effect the light curve. Compact winds can be analyzed using the diffusion approximation, but extended winds require a more accurate treatment of radiation transport.

We calculated light curves, and especially light curve timescales, for both compact and extended winds using a simple stationary configuration, solved by a Monte Carlo method. We focus on winds with a density profile of $\rho(r)=K r^{-2}$, which would be the result of a steady mass loss from the progenitor star. Our results for compact winds are in agreement with previous analytical and numerical diffusion analysis. The results for extended winds, on the other hand, deviate from diffusion estimates. We show that the figure of merit in this context is the ratio $\tilde R_w\equiv R_w/(\kappa K)$, where $R_w$ is the wind outer radius, and $\kappa$ is the specific opacity ($\kappa K$ is essentially the typical optical length scale in the wind). We find that for $\tilde R_w \gtrsim 1$, light curves calculated with the diffusion approximation overestimate the peak luminosity and underestimate the timescales (both full width at half maximum and rise time) of the light curve by a few tens of percents. Hence, while the diffusion approximation does allow to relate the properties of the progenitor system to the observed light curve in terms of the correct order of magnitude (for $\tilde R_w<10^6$), it may not suffice when accurate modeling is required (see Figure \ref{fig:time_scale_semi}).

Since we base our analysis on a schematic stationary model, we do not attempt to reconstruct actual observed light curves. We can, however, asses the relevance of our results to different sub-classes of interacting supernovae. In the case of superluminous supernovae (SLSNe), the total mass in the wind must be comparable to the mass in the ejecta for an efficient conversion of kinetic energy to radiation (see Section \ref{sec:timescale}). For wind velocities of order $10^6\textrm{ cm s}^{-1}$, a mass loss period of a few hundred years will place the wind edge at a few $10^{15}$ cm. If this wind is to hold a mass of order $10 M_\sun$, it must have $K\sim 10^{18}\textrm{ g cm}^{-1}$, yielding $\kappa K\sim 10^{17}$ cm, and so $\tilde R_w\sim 10^{-2}$ (assuming $\kappa\approx 0.34\textrm{ cm}^2\textrm{g}^{-1}$). Correspondingly, SLSNe with an inferred photospheric radius of a few $10^{15}$ cm, such as those we considered in \citet{GB2012}, can be analyzed to good precision with the diffusion approximation.

The situation may be different in one of two cases. First, still in the context of SLSNe, a massive wind which extends to larger radii can lead to an $\tilde R_w\approx 1$ situation (see, for example, the $2.8M_\sun$ wind in Figure \ref{fig:light_curves_ext}). Such a system can arise either by a faster wind, for example, driven from a more compact progenitor, such as a Wolf-Rayet star, or if the period of mass loss extends over longer timescales, of thousands of years. Either way, if $R_w$ is somehow a few $10^{16}$ cm, a $10M_\sun$ wind will correspond to $\tilde R_w\approx 1$, and corrections to a diffusion approximation will be in order.

The second case of interest is in Type IIn supernovae which are not superluminous, presumably if the wind mass is much smaller than the ejecta mass. If the wind mass is only about $0.1M_\sun$ or less, a value of $\tilde R_w\approx 1$ may arise even for slow winds, which extend to a few $10^{15}$ cm.
Some recent examples of such supernovae are SN 2005gl \citep{GalyamLeonard2009} and SN (candidate) 2009ip \citep{Margutti2013,Ofek2013,Smith2013}. Some of the synthetic models considered by \citet{Moriya2011,Moriya2013,Chatzopoulos2013} also correspond to this regime. We caution that in very light winds the resulting emission during shock breakout may no longer dominate the light curve, but rather generate only a short precursor. Our stationary model is, however, a weaker approximation to this case, since the velocities during breakout through a dilute wind can be quite high.

We compared the effect of deviation from diffusive radiation flow to that of light travel time difference, of photons originating from different positions on the surface of the star. The latter effect also modulates the light curve measured by a distant observer. We showed that light travel time differences can be completely neglected for compact winds (but not for bare stars, which are essentially extremely compact winds; see Section \ref{sec:observer} and Figure \ref{fig:travel}), but should be taken into account for extended winds. Nevertheless, after including the light travel time difference in the Monte Carlo calculations, this effect was found to influence the light curve timescales by no more than 10\% (see Figure \ref{fig:travel_semi}). The relative secondary significance of this effect is not trivial, and could not be appreciated without Monte Carlo calculations, or a complete analytical solution.

A complimentary conclusion of this work regards the widely used flux-limited diffusion approximation. This method has been suggested \citep[see, e.g.,][]{LP81,TurnerStone,Castor,Frey2012} as a computationally cheap approximation of photon transfer, which is more accurate than simple diffusion. However, as we show in Section \ref{subsec:flux_limited_lc} and in Figure \ref{fig:flux_limit}, flux-limited diffusion gives qualitatively wrong results for extended winds. Since simple diffusion is sufficient for describing compact winds, and flux-limited diffusion does not reproduce the Monte Carlo results for extended winds, it seems that the relevancy of this method in the context of light curve calculations requires reassessment. For reliable light curves in the extended wind regime, full transport or Monte Carlo methods are required.

\acknowledgements 

We are grateful to Eli Livne for helpful discussions.

\appendix
\section{Flux Limiter Approach}
\label{sec:flux_limiter}

The diffusion approximation may be written as
\begin{equation}\label{eq:diff_approx}
F=-D\nabla U,
\end{equation}
with $F$ denoting the radiative flux and $U$ the radiation energy density. $D$ is the diffusion coefficient, which is equal to
\begin{equation}\label{eq:diff_coeff}
D=\frac{lc}{3},
\end{equation}
with $l=1/\kappa\rho$ denoting, as before, the mean free path. The derivation of this approximation is based upon an assumption of weak anisotropy \citep[see][for a complete derivation]{Zeldovich}. The weak anisotropy assumption, and therefore the diffusion approximation, breaks down at low optical depth. At the extreme optically thin limit, and assuming an appropriate symmetry, all the photons travel in the same direction (for example, outward), inducing a flux
\begin{equation}\label{eq:flux_stream}
|F|=cU.
\end{equation}
This case is refereed to as the ``free-streaming'' limit, and it is the case of maximum anisotropy.

The transition between the optically thick, diffusion, limit and the optically thin, free-streaming, limit occurs naturally in Monte Carlo methods, which take into account the angular distribution of the photons. Alternatively, the angular information may be approximated. One such approximation method is the widely used flux-limited diffusion. We follow by presenting in Section \ref{subsec:flux_limited_theory} the flux-limited diffusion approximation, as described by \citet{LP81, Levermore1983} and implemented by \citet{TurnerStone}. 
In Section \ref{subsec:flux_limited_lc} we use this approximation in calculations of the light curve of the supernova-wind scenario, and compare the results with those presented above for diffusion and Monte Carlo simulations. In Section \ref{subsec:free_streaming} we demonstrate the application of the approximation to the free-streaming limit and discuss its limitations.

\subsection{Flux-Limited Diffusion Theory}
\label{subsec:flux_limited_theory} 

In flux-limited diffusion theory, the radiative flux is calculated using Equation \eqref{eq:diff_approx}, but the diffusion coefficient takes the form
\begin{equation}\label{eq:flux_limiter_d}
D=\lambda lc,
\end{equation}
with $\lambda$ is an introduced dimensionless quantity called the flux-limiter. $\lambda$ is a function of another dimensionless quantity
\begin{equation}\label{eq:flux_r}
R=\frac{|\nabla U|l}{U}.
\end{equation}
Intuitively, $R$ can be related to the optical depth $\tau$, by estimating $\nabla U\approx U/x$, with $x$ is of the order of the distance over which $U$ changes \citep[the distance from the edge in our case; see][for a similar argument]{Zeldovich}. This gives the estimate $R\approx l/x=1/\tau$. With this identification, we can formulate the limits which the function $\lambda(R)$ should comply with:
\begin{equation}\label{eq:lambda_limits}
\lambda(R)=
\begin{cases}
1/3 & R\to 0
\\
1/R & R\to\infty
\end{cases}.
\end{equation}
Thus, at $\tau\to\infty$, or $R\to 0$, Equation \eqref{eq:flux_limiter_d} reduces to the original diffusion approximation coefficient of Equation \eqref{eq:diff_coeff}, and for $\tau\to 0$, or $R\to\infty$, Equations \eqref{eq:diff_approx}, \eqref{eq:flux_limiter_d}, and \eqref{eq:flux_r} give the free-streaming limit of Equation \eqref{eq:flux_stream}. The interpolation of $\lambda(R)$ between the two limits depends on the details of the angular distribution of the photons. Since this distribution is not known (without a transport/ Monte Carlo calculation), the choice of flux limiter is arbitrary, as long as it complies with Equation \eqref{eq:lambda_limits}. In the current work, we choose the flux limiter of \citet{LP81}, which is obtained by applying certain assumptions on the angular distribution. The flux limiter takes the form 
\begin{equation}\label{eq:LP_full}
\lambda(R)=\frac{1}{R}\left(\coth R-\frac{1}{R}\right).
\end{equation}
For numerical reasons, we use a rational approximation to Equation \eqref{eq:LP_full}, which was derived by \citet{LP81}
\begin{equation}\label{eq:LP_approx}
\lambda(R)=\frac{2+R}{6+3R+R^2}.
\end{equation}

We incorporate the flux limiter, as given by Equations \eqref{eq:flux_limiter_d}, \eqref{eq:flux_r}, and \eqref{eq:LP_approx} in the stationary diffusion code. In addition, we set the appropriate free-streaming boundary condition, $F=cU$ at the edge.

\subsection{Flux-Limited Light Curves}
\label{subsec:flux_limited_lc}

\begin{figure}[tbh]
\epsscale{0.5} \plotone{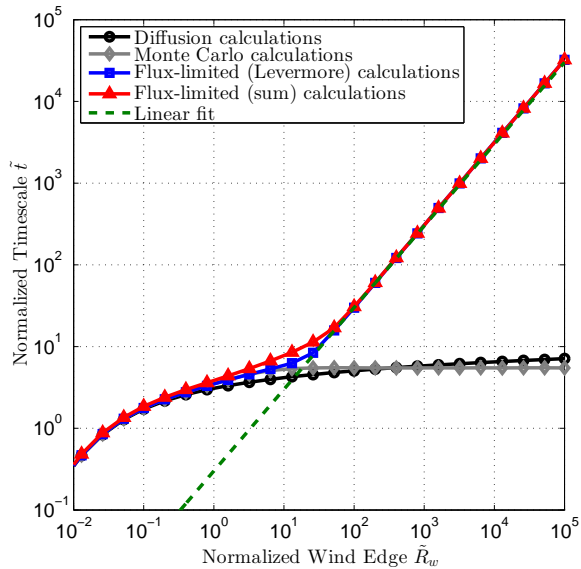}
\caption{Calculated normalized (see text) light curve timescales $\tilde t$ (full width at half maximum) as a function of the normalized wind outer radius $\tilde R_w$. The results are plotted for $\tilde R_d=10^{-2}$. The light curves are calculated using the diffusion method without a flux limiter (solid black line, marked with circles), with the Levermore flux limiter (solid blue line, marked with squares), and with the sum flux limiter (solid red line, marked with triangles). For comparison, Monte Carlo calculations (solid grey line, marked with diamonds) are added. Each marker represents a single simulation. The analytic fit (dashed green line) is linear in $\tilde R_w$.
\label{fig:flux_limit}}
\end{figure}

We repeated the diffusion calculations of Section \ref{sec:light_curves} with a flux limiter. The results are presented in Figure \ref{fig:flux_limit}. For compact winds ($\tilde R_w\ll 1$), the flux-limited and the simple diffusion calculations in Figure \ref{fig:flux_limit} coincide, as expected from the discussion in Section \ref{sec:timescale}. These results also comply with the Monte Carlo calculations, as seen in Figure \ref{fig:flux_limit}.

We now turn our focus to the extended region ($\tilde R_w\gtrsim 1$). In this regime, the flux-limited results separate from the simple diffusion calculations, but fail to reproduce the correct behavior, found using the Monte Carlo method.

The Monte Carlo calculations suggest that the timescale tends to an asymptotic constant in the ultra extended limit ($\tilde R_w\gg 1$). The flux-limited timescales, on the other hand, exhibit a linear growth with $\tilde R_w$. This linear behavior is not a feature of the particular flux limiter we chose. As seen in Figure \ref{fig:flux_limit}, we obtain similar results when using a simpler, ``sum'' flux limiter,
\begin{equation}
\lambda(R)=\frac{1}{3+R},
\end{equation} 
instead of the Levermore flux limiter of Equation \eqref{eq:LP_approx}. It is worth noting that we obtain similar results when incorporating a flux limiter in the hydro-diffusion code described in \citet{GB2012}. Such an incorporation in a hydrodynamic code requires an additional change of the radiation pressure \citep[see][]{Levermore1983,TurnerStone}.

This qualitatively wrong behavior is not surprising, since flux limiters were originally introduced as an ad hoc solution for the violation of causality in diffusion calculations. Despite attempts to introduce physically more justified, and less arbitrary limiters \citep{LP81}, flux-limited diffusion remains only an approximation which does not capture the angular distribution exactly \citep[see][for similar conclusions]{TurnerStone}. In the particular problem discussed in this work, it seems that flux-limited diffusion yields results which are not only quantitatively, but also qualitatively wrong. The origin of this discrepancy is the diffusive nature of the difference equations, even at the free-streaming limit, as discussed in Section \ref{subsec:free_streaming}. 
Flux-limited diffusion is not completely pointless, and captures some of the physics in this scenario. For example, the light curve emergence is delayed by the correct value of roughly $R_w/c$, an effect completely missed by simple diffusion calculations (see Section \ref{subsec:free_streaming}). However, since the light curve timescale is related to the variance in photon arrival times, and not to the average arrival time, flux-limited diffusion leads to incorrect results.

\subsection{Free-Streaming Photon Transfer}
\label{subsec:free_streaming}

It is evident from Figure \ref{fig:flux_limit} that the flux-limited diffusion approximation deviates considerably from the Monte Carlo results in the ultra extended regime ($\tilde R_w\gg 1$). This regime corresponds to winds which are dominated by very optically thin regions, and therefore are unaffected by the details of the transition from diffusive (optically thick) to free-streaming (optically thin) photon transfer (for example, different flux limiters converge, as seen in Figure \ref{fig:flux_limit}). In this regime, the photons should free-stream at the speed of light, without widening the light curve.

In this section we reproduce the results of Section \ref{subsec:flux_limited_lc} in a simple free-streaming scenario, and thereby demonstrate the origin of the discrepancy between the flux-limited and Monte Carlo results.

\begin{figure}[tbh]
\epsscale{0.5} \plotone{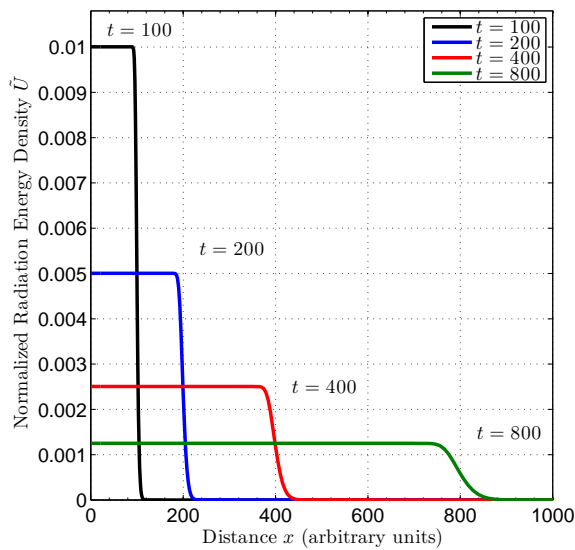}
\caption{Propagation of a planar flux-limited radiation wave in a completely transparent medium. The normalized radiation energy density $\tilde U$ as a function of the distance from the origin $x$ is plotted at different times following the energy release ($t=100,200,400,800$ as solid black, blue, red, and green lines respectively). The distance and time units are arbitrary, but connected through the choice $c=1$.
\label{fig:flux_limit_wave}}
\end{figure}

In a similar fashion to Section \ref{sec:deviations}, we release a unit energy at the innermost cell of a stationary configuration. In this section we use a very simple configuration of a semi-infinite planar slab, which is practically transparent. This configuration is achieved numerically by setting a slab of finite length with an infinitesimal density, so the total optical depth is $\sim 10^{-7}$ (the results are not sensitive to the optical depth as long as it is infinitesimally small), and by examining times before the radiation wave reaches the edge. In Figure \ref{fig:flux_limit_wave} we follow the propagation of the resulting radiation wave.

\begin{figure}[tbh]
\epsscale{0.5} \plotone{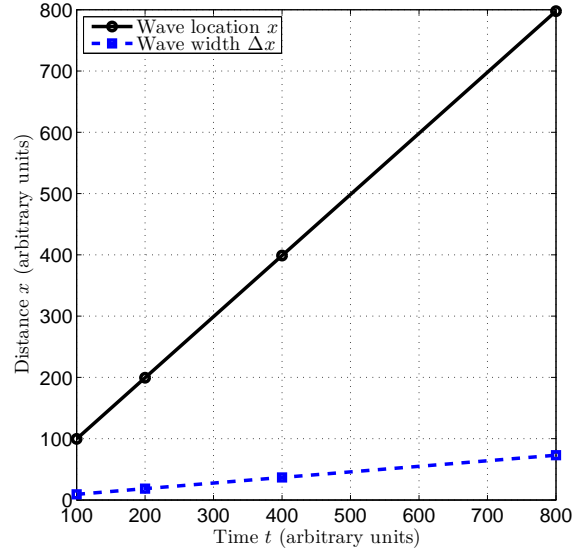}
\caption{Propagation of a planar flux-limited radiation wave in a completely transparent medium. The wave location and width are determined from Figure \ref{fig:flux_limit_wave}. The wave location $x$ (solid black line, marked with circles) is determined by the rise to 50\% of the maximum radiation energy density, and the wave width $\Delta x$ (dashed blue line, marked with squares) is determined by the rise distance from 10\% to 90\% of peak radiation energy density. Each marker corresponds to a different time $t$ (see Figure \ref{fig:flux_limit_wave}). The distance and time units are arbitrary, but connected through the choice $c=1$.
\label{fig:flux_limit_width}}
\end{figure}

Since the medium is transparent, the expected solution is an ideal step function
\begin{equation}\label{eq:step}
\tilde U\left(x,t\right)=\frac{1}{ct}\Theta\left(ct-x\right),
\end{equation}  
traveling at the speed of light, where $t$ is the time and $x$ is the distance. For simplicity, we set the speed of light $c=1$. However, as seen in Figure \ref{fig:flux_limit_wave}, the flux-limited calculation smears the ideal step function. Moreover, the smearing increases as the radiation wave propagates. In Figure \ref{fig:flux_limit_width} we present the wave location and width as a function of time. This figure demonstrates how the flux limiter does succeed in its main goal of creating a radiation wave which travels at the speed of light. This is its main advantage over the simple diffusion approximation, which would result in an instantaneous release of the energy at an infinite speed. However, as seen in Figure \ref{fig:flux_limit_width}, the wave smearing increases linearly with time (and distance). This spatial widening of the wave front is the cause for the light curve widening as the wave reaches the wind's edge, and the origin of the linear asymptotic behavior in Figure \ref{fig:flux_limit}. The widening of the wave front is due to the diffusive nature of the difference equations, which do not converge to the non-diffusive wave equation. In addition to the smearing of the wave front, which causes the widening of the light curve rise time, the overall dissipation of the radiation energy over a distance of $ct$ is the cause for the widening of the decline time. This behavior is also due to the diffusive nature of the difference equations, which impose a flux in the opposite direction of $\nabla U$, in contrast to free streaming. 

\bibliographystyle{apj}

\end{document}